\documentclass{elsart}
\newcommand{\eo}[2]{\left\{\begin{array}{c}#1\\[-2mm]#2\end{array}\right\}}

\begin{document}
\begin{frontmatter}
\title{Summing next-to-next-to-leading logarithms\\
in $b \to c$ transitions at zero recoil}
\author{A.G.~Grozin}
\ead{grozin@particle.uni-karlsruhe.de}
\address{Institut f\"ur Theoretische Teilchenphysik,
Universit\"at Karlsruhe}
\begin{abstract}
Perturbative corrections to $b\to c$ transitions at zero recoil
are considered in the two-step matching scheme.
The matching coefficient for the $b\to c$ currents
from the intermediate effective theory (between the scales $m_b$ and $m_c$)
to the low-energy effective theory (scales below $m_c$)
has been found with two-loop accuracy.
The next-to-next-to-leading logarithms has been summed
in the leading order in $m_c/m_b$.
Higher-order corrections are estimated in the large-$\beta_0$ limit.
\end{abstract}
\end{frontmatter}

\section{Introduction}
\label{S:Intro}

Precise measurement of $V_{cb}$ is one of important tasks
in the investigation of the CKM mixing matrix~\cite{CKM}.
Two methods are currently used: inclusite and exclusive.
Here we shall discuss the exclusive method.

Weak $b\to c$ transitions are described
by QCD currents $\bar{c}\Gamma b$
with $\Gamma=\gamma^\mu$ and $\gamma_5 \gamma^\mu$.
If we are only interested in their matrix elements
for $p_b=m_b v+k$, $p_c=m_c v'+k'$,
where the residual momenta $k$, $k'$
are small as compared to the heavy-quark masses $m_{c,b}$,
then we may expand these QCD currents in HQET operators
of the form $\bar{c}_{v'}\cdots b_v$
with the appropriate quantum numbers
divided by powers of $m_{c,b}$.
The coefficients are calculated by matching.
This is descussed, e.g., in the review~\cite{Ne:94}
or in the textbook~\cite{G:04}.
In this paper we consider the zero-recoil case $v'=v$.
Then, in the $v$ rest frame,
\begin{eqnarray}
\bar{c} \gamma_0 b &=& \eta_V\,\bar{c}_v b_v
+ \mathcal{O}(1/m_{c,b}^2)\,,
\nonumber\\
\bar{c} \gamma_5 \vec{\gamma} b &=& \eta_A\,
\bar{c}_v \gamma_5 \vec{\gamma} b_v
+ \mathcal{O}(1/m_{c,b}^2)\,.
\label{Intro:def}
\end{eqnarray}
There are no $1/m_{c,b}$ corrections~\cite{L:90}.
We use $\overline{\mathrm{MS}}$ renormalization
with anticommuting $\gamma_5$;
then the vector and axial currents in QCD do not depend
on the normalization scale $\mu$.
The HQET currents with $v'=v$ are also $\mu$-independent,
and hence the matching coefficients $\eta_{V,A}$
do not depend on any normalization scale.

In the exclusive method, the differential rate of
$\bar{B}\to D^* l\bar{\nu}$ is extrapolated
to the zero-recoil point.
Thus, $F(1)|V_{cb}|$ is measured,
where $F(1)=\eta_A[1+\mathcal{O}(1/m_{c,b}^2)]$.
Matrix elements of higher-dimensional HQET operators
in the power correction are estimated in several ways,
giving about 4\% error in $F(1)$~\cite{CKM}.
This error can be reduced to 1\% level
by unquenched lattice simulations currently in progress.
Therefore, it is desirable to know $\eta_\Gamma$
with the accuracy better than 1\%.

These coefficients can be obtained by matching QCD matrix elements
with those in HQET:
\begin{equation}
\eta_\Gamma = 1 + \sum_n \eta_\Gamma^{(n)}
\left(\frac{\alpha_s^{(5)}(\mu_0)}{4\pi}\right)^n\,,\quad
\mu_0 = \sqrt{m_b m_c}\,.
\label{Intro:1step}
\end{equation}
Here $m_{b,c}$ are the on-shell heavy-quark masses.
In this way, we know the exact dependence of $\eta_\Gamma^{(n)}$
on $r=m_c/m_b$;
however, we cannot sum terms with leading powers of $L=\log(m_b/m_c)$
in all orders of perturbation theory.
The two-loop correction $\eta_\Gamma^{(2)}$
has been calculated in~\cite{CM:97}
and confirmed by an independent re-calculation~\cite{FT:98}.
The three-loop correction has been recently calculated~\cite{AC:04}
in the equal-mass limit $m_b=m_c$.

Terms with the highest powers of $\beta_0$
in all orders of the perturbative series
were summed in~\cite{Ne:95}.
The leading infrared renormalon
in zero-recoil matching coefficients is at $u=1$.
This means that high-order behaviour of perturbation series
for $\eta_\Gamma$ is much better than in, e.g.,
the pole mass (where it is at $u=1/2$).
We don't have to worry about it yet.

In this single-step matching approach,
the evolution of $\alpha_s$ and the relevant operators
between the scales $m_b$ and $m_c$
is not taken into account properly.
In this interval, there are 4 active flavours;
$b$ is heavy while $c$ is still light,
and the evolution of operators is governed by HQET
(in particular, $\alpha_s^{(5)}(\mu_0)$ is not a natural quantity).
And with $\alpha_s^{(4)}(m_c)/\alpha_s^{(4)}(m_b)\approx1.56$,
these evolution effects are of order 1, not small corrections.
Therefore, it is natural to take them into account
from the outset.

This is achieved by performing the matching in two steps.
First, we match QCD with HQET-1,
where $b$ is static and $c$ is dynamic, at $\mu\sim m_b$.
Second, we match HQET-1 with HQET-2,
where both $b$ and $c$ are static, at $\mu\sim m_c$.
In this way, we obtain $\eta_{V,A}$ as series in $r$:
\begin{equation}
\eta_\Gamma = C_\Gamma(m_b) G(m_b,m_c) E(m_c) + \mathcal{O}(r)\,.
\label{Intro:2step}
\end{equation}
In practice, few terms of this expansion can be found,
because the number of relevant operators in HQET-1
grows fast with the dimensionality.
However, in each term of the expansion in $r$,
we can sum leading, next-to-leading, etc., powers of $L$.
The leading term in the $r$-expansion is known
at the next-to-leading logarithmic order~\cite{Ne:92}.
The $\mathcal{O}(r)$ term is known
at the leading logarithmic order~\cite{FG:90}.
The $\mathcal{O}(r^2)$ term is discussed in~\cite{BO:97}.
The two approaches can be combined:
we subtract several first terms of expansion of~(\ref{Intro:1step})
in $r$ and add the RG-improved results instead.

The QCD $\to$ HQET-1 matching coefficients
\begin{equation}
C_\Gamma(m_b) = 1 + \sum_n c_\Gamma^{(n)}
\left(\frac{\alpha_s^{(5)}(m_b)}{4\pi}\right)^n
\label{Intro:C}
\end{equation}
are known at the next-to-next-to-leading (NNL) order~\cite{BG:95,G:97}.
The running factor $G(m_b,m_c)$
is determined by the anomalous dimension $\gamma$
of the heavy-light current in HQET-1.
Until recently, it was only known at two loops~\cite{JM:91,BG:91}.
Now it is has been calculated at three loops~\cite{CG:03},
i.e., also at the NNL order,
using the methods of~\cite{G:00}.
Finally, the HQET-1 $\to$ HQET-2 matching coefficient
\begin{equation}
E(m_c) = 1 + \sum_n e_n
\left(\frac{\alpha_s^{(4)}(m_c)}{4\pi}\right)^n
\label{Intro:E}
\end{equation}
is currently known at one loop~\cite{Ne:92} only.
In the present paper,
we shall find the two-loop coefficient $e_2$,
thus completing the NNL order calculation
of $\eta_{V,A}$ at $\mathcal{O}(r^0)$.

\section{Summing NNL logarithms}
\label{S:NNL}

By requiring that the two-step matching result~(\ref{Intro:2step})
(with the three-loop $\gamma$~\cite{CG:03})
reproduces the $\mathcal{O}(r^0)$ term in $\eta_{V,A}^{(2)}$~\cite{CM:97},
we have two independent derivations of the matching coefficient $E(m_c)$:
\begin{eqnarray}
E(m_c) &=& 1 - 4 C_F \frac{\alpha_s(m_c)}{4\pi}
\nonumber\\
&&{} + C_F \Biggl[ C_F \left( - 20 \pi^2 \log 2 + 14 \zeta_3 + 73 \zeta_2
+ \frac{81}{16} \right)
\nonumber\\
&&\hphantom{{}+C_F\Biggl[\Biggr.}
+ C_A \left( 10 \pi^2 \log 2 - 41 \zeta_3 - 6 \zeta_2 + \frac{503}{48}
\right)
\nonumber\\
&&\hphantom{{}+C_F\Biggl[\Biggr.}
+ T_F \left( - 17 \zeta_2 + \frac{233}{12} \right)
+ T_F n_l \left( - 4 \zeta_2 - \frac{31}{12} \right) \Biggr]
\left(\frac{\alpha_s(m_c)}{4\pi}\right)^2
\nonumber\\
&\approx& 1 - \frac{4}{3} \frac{\alpha_s(m_c)}{\pi}
+ 4.00 \left(\frac{\alpha_s(m_c)}{\pi}\right)^2\,.
\label{NNL:E}
\end{eqnarray}
Here $n_l=3$ is the number of flavours lighter than $c$;
$\alpha_s^{(4)}(\mu)$ is written simply as $\alpha_s(\mu)$.
This result can also be obtained by a direct calculation;
this will be discussed in a separate publication.

Now we can sum the logarithms (up to NNL ones) to all orders:
\begin{equation}
\eta_\Gamma = x^{-\gamma_0/(2\beta_0)}
\left[ 1 + h_\Gamma^{(1)} \frac{\alpha_s(m_b)}{4\pi}
+ h_\Gamma^{(2)} \left(\frac{\alpha_s(m_b)}{4\pi}\right)^2
+ \cdots \right]\,.
\label{NNL:logs}
\end{equation}
Here
\begin{equation}
x = \frac{\alpha_s(m_c)}{\alpha_s(m_b)}\,,
\label{NNL:x}
\end{equation}
$\alpha_s(\mu)$ and $\beta_0$ are for 4 flavours.
The NL correction is
\begin{eqnarray}
\frac{h_{V,A}^{(1)}}{C_F} &=&
- \frac{1}{4} \left( 11 x + \eo{13}{21} \right)
+ \left[ \left( 8 \zeta_2 - \frac{23}{4} \right) C_F
- 2 ( \zeta_2 + 4 ) C_A \right]
\frac{x-1}{\beta_0}
\nonumber\\
&&{} + \frac{3 C_A (11 C_F + 7 C_A) (x-1)}{2 \beta_0^2}\,,
\label{NNL:NL}
\end{eqnarray}
where the upper number in braces is for $V$
and the lower one for $A$.
The NNL correction is
\begin{eqnarray}
&&\frac{h_{V,A}^{(2)}}{C_F} =
\left[ \left( 3 \zeta_2 + \frac{29}{24} \right) x^2
- 3 \zeta_2 - \frac{1}{24} \eo{53}{205} \right] \beta_0
\nonumber\\
&&{} + \left[ \left( - 20 \pi^2 \log 2 + 3 \zeta_3 + \frac{247}{3} \zeta_2
+ \frac{643}{96} \right) x^2
+ \eo{143}{231} \frac{x}{16} \right.
\nonumber\\
&&\quad\left.{} + \eo{4}{28/3} \pi^2 \log 2 - \eo{3}{11} \zeta_3
- \eo{73/3}{67} \zeta_2 + \frac{1}{96} \eo{1283}{2899} \right] C_F
\nonumber\\
&&{} + \left[ \left( 10 \pi^2 \log 2 - \frac{63}{2} \zeta_3
- \frac{58}{3} \zeta_2 - \frac{41}{96} \right) x^2
\vphantom{\eo{1}{1}} \right.
\nonumber\\
&&\quad\left.{} - \eo{2}{14/3} \pi^2 \log 2 - \eo{9}{1} \frac{\zeta_3}{2}
+ \eo{55}{103} \frac{\zeta_2}{3} - \frac{1}{96} \eo{343}{919} \right] C_A
\nonumber\\
&&{} + \left[ - (13 \zeta_2 - 22) x^2
- \eo{24}{40/3} \zeta_2 + \eo{727/18}{133/6} \right] T_F
\nonumber\\
&&{} + \left\{ \left[ \left( 20 \zeta_4 + 9 \zeta_3 - 50 \zeta_2
+ \frac{483}{16} \right) x
\vphantom{\eo{1}{1}} \right. \right.
\nonumber\\
&&\qquad\left.{} + 20 \zeta_4 + 9 \zeta_3 - \eo{54}{70} \zeta_2
+ \frac{1}{16} \eo{529}{713} \right] C_F^2
\nonumber\\
&&\quad{} + \left[ \left( 4 \zeta_4 + \frac{57}{2} \zeta_3
- \frac{77}{6} \zeta_2 + \frac{629}{24} \right) x
\vphantom{\eo{1}{1}} \right.
\nonumber\\
&&\qquad\left.{} + 4 \zeta_4 + \frac{57}{2} \zeta_3
- \eo{71}{47} \frac{\zeta_2}{6} + \frac{1}{24} \eo{725}{1109} \right] C_F C_A
\nonumber\\
&&\quad\left.{} + \left( 6 \zeta_4 - 33 \zeta_3 + \frac{19}{3} \zeta_2
+ \frac{103}{8} \right) (x+1) C_A^2
\vphantom{\eo{1}{1}} \right\} \frac{x-1}{\beta_0}
\nonumber\\
&&{} + \left\{ \frac{1}{32} (32 \zeta_2 - 23)^2 (x-1) C_F^3
- \left( 7 \zeta_2 + \frac{609}{32} \right) (x+1) C_A^3
\vphantom{\eo{1}{1}} \right.
\nonumber\\
&&\quad{} + \left[ - \left( 40 \zeta_4 + \frac{17}{2} \zeta_2
+ \frac{479}{8} \right) x + 40 \zeta_4 + \frac{193}{2} \zeta_2
- \frac{1}{8} \eo{1281}{1545} \right] C_F^2 C_A
\nonumber\\
&&\quad\left.{} + \left[ \left( 5 \zeta_4 + 33 \zeta_2
- \frac{395}{4} \right) x
- 5 \zeta_4 + \zeta_2 - \eo{168}{189} \right] C_F C_A^2
\right\} \frac{x-1}{\beta_0^2}
\nonumber\\
&&{} + 3 C_A (11 C_F + 7 C_A) \left\{
\left( 4 \zeta_2 - \frac{23}{8} \right) (x-1) C_F^2
+ \frac{7}{4} (x+1) C_A^2 \right.
\nonumber\\
&&\quad\left.{} + \left[ - \left( \zeta_2 + \frac{5}{4} \right) x
+ \zeta_2 + \frac{27}{4} \right] C_F C_A \right\}
\frac{x-1}{\beta_0^3}
\nonumber\\
&&{} + \frac{9 C_F C_A^2 (11 C_F + 7 C_A)^2 (x-1)^2}{8 \beta_0^4}\,.
\label{NNL:NNL}
\end{eqnarray}
Numerically,
\begin{eqnarray}
&&\eta_{V,A} = x^{6/25} \left[ 1
- \left( 1.10588 x + \eo{0.89412}{1.56079} \right)
\frac{\alpha_s(m_b)}{\pi} \right.
\nonumber\\
&&\left.{} + \left( 3.94156 x^2 + \eo{0.98879}{1.72604} x
- \eo{2.44780}{8.22155} \right)
\left(\frac{\alpha_s(m_b)}{\pi}\right)^2
+ \cdots \right]\,.
\label{NNL:num}
\end{eqnarray}

\section{Higher orders in the large-$\beta_0$ limit}
\label{S:beta}

It would be extremely difficult to sum the N$^3$L logarithms.
However, it is easy to find the $\beta_0^2$ term in $h_\Gamma^{(3)}$.
The highest powers of $\beta_0$ in all orders of perturbation theory
for $\gamma$ were found in~\cite{BG:95}:
\begin{eqnarray}
\gamma &=& - \frac{C_F}{3} \frac{\beta}{\beta_0}
\frac{3+2\beta}{B(2+\beta,2+\beta)\Gamma(1-\beta)\Gamma(3+\beta)}
\nonumber\\
&=& - 3 C_F \frac{\beta}{\beta_0} \left[ 1 + \frac{5}{6} \beta
- \frac{35}{36} \beta^2 - \left( 2 \zeta_3 - \frac{83}{72} \right) \beta^3
+ \cdots \right]\,.
\label{NNL:gamma}
\end{eqnarray}
where $\beta=\beta_0\alpha_s/(4\pi)$.
The matching coefficients $C_\Gamma(m_b)$
were also considered in~\cite{BG:95}:
\begin{eqnarray}
C_{V,A}(m_b) &=& 1 - C_F \frac{\beta}{\beta_0}
\left[ \eo{2}{4} + \left( 3 \zeta_2 + \eo{47/16}{445/48} \right) \beta
\right.
\nonumber\\
&&\left.{} + \left( 7 \zeta_3 + \eo{13}{21} \zeta_2
+ \frac{1}{432} \eo{-1751}{7993} \right) \beta^2
+ \cdots \right]\,.
\label{NNL:C}
\end{eqnarray}

Here we use the same method for $E$.
The solution of the RG equation can be written as
\begin{eqnarray}
&&E(\mu) = \hat{E}
\left(\frac{\alpha_s(\mu)}{\alpha_s(m_c)}\right)^{\gamma_0/(2\beta_0)}
K(\alpha_s(\mu))\,,
\nonumber\\
&&K(\alpha_s) = \exp \int_0^{\alpha_s}
\left(\frac{\gamma(\alpha_s)}{2\beta(\alpha_s)}
-\frac{\gamma_0}{2\beta_0}\right)
\frac{d\alpha_s}{\alpha_s}\,.
\label{NNL:K}
\end{eqnarray}
The RG invariant $\hat{E}$ is,
at the first order in $1/\beta_0$,
\begin{equation}
\hat{E} = 1 + \frac{1}{\beta_0} \int_0^\infty
S(u) e^{-u/\beta} du
+ \mathcal{O}\left(\frac{1}{\beta_0^2}\right)\,,
\label{NNL:Ehat}
\end{equation}
where $\beta=\beta_0\alpha_s(m_c)/(4\pi)$.
We obtain
\begin{equation}
S(u) = 3 C_F \left[ e^{(5/3)u}
\frac{\Gamma(u)\Gamma(2-2u)}{\Gamma(3-u)}
\frac{1-u-u^2}{1+2u} - \frac{1}{2u} \right]\,.
\label{NNL:Su}
\end{equation}
Note that the leading infrared renormalon is at $u=1$.
Collecting all this together, we obtain
\begin{eqnarray}
E(m_c) &=& 1 + C_F \frac{\beta}{\beta_0} \left[ - 4
+ \left( 3 \zeta_2 + \frac{31}{16} \right) \beta \right.
\nonumber\\
&&\left.{}
+ \left( 7 \zeta_3 - 11 \zeta_2 - \frac{7655}{432} \right) \beta^2
+ \cdots \right]\,.
\label{NNL:E3}
\end{eqnarray}
This reproduces the $\beta_0$ term at two loops~(\ref{NNL:E}),
and gives us the $\beta_0^2$ term at three loops.

Using this result, we obtain
\begin{eqnarray}
\frac{h_{V,A}^{(3)}}{C_F} &=& \left[
\left( 6 \zeta_3 - 11 \zeta_2 - \frac{3703}{216} \right) x^3
- 6 \zeta_3 - \eo{13}{21} \zeta_2 + \frac{1}{216} \eo{751}{-4121} \right]
\beta_0^2
\nonumber\\
&&{} + \mathcal{O}(\beta_0)\,.
\label{NNL:NNNL}
\end{eqnarray}
Numerically, this means adding the term
\[
- \left( 40.5461 x^3 + \eo{36.3421}{88.0131} \right)
\left(\frac{\alpha_s(m_b)}{\pi}\right)^3
\]
inside the bracket in~(\ref{NNL:num}).
Of course, this is an estimate only,
because the terms $\mathcal{O}(\beta_0)$ and $\mathcal{O}(1)$
are not known.

\section{What is known about the three-loop coefficients?}
\label{S:3}

Re-expressing $\alpha_s(m_b)$, $\alpha_s(m_c)$ via $\alpha_s^{(5)}(\mu_0)$,
we can write $\eta_\Gamma$ in the form~(\ref{Intro:1step}).
From our NNL result, we obtain the $L^3$, $L^2$, and $L$ terms
in the three-loop coefficients $\eta_\Gamma^{(3)}$.
The $\beta_0^2$ part of the non-logarithmic term can also be found.
We can even make one more improvement,
and obtain the $\beta_0^2$ part of $\eta_\Gamma^{(3)}$
exactly as a function of $r$ (including power corrections).
The highest powers of $\beta_0$ in all orders of perturbation theory
for $\eta_\Gamma$ were found in~\cite{Ne:95}.
They have the form similar to~(\ref{NNL:Ehat}),
with $\beta=\beta_0\alpha_s(\mu_0)/(4\pi)$ and
\begin{eqnarray}
S_V(u) &=& 6 C_F e^{(5/3)u}
\frac{\Gamma(u)\Gamma(1-2u)}{\Gamma(3-u)}
\frac{1-u-u^2}{1+2u}
\left[ (1-2u) R_0 - R_1 \right]\,,
\nonumber\\
S_A(u) &=& 2 C_F e^{(5/3)u}
\frac{\Gamma(u)\Gamma(1-2u)}{\Gamma(3-u)}
\frac{1}{1+2u}
\nonumber\\
&&{}\times
\left[ 3 (1-2u) (1-u-u^2) R_0 - (3-u+u^2) R_1 \right]\,,
\label{3:Su}
\end{eqnarray}
where
\begin{equation}
R_0 = \cosh(Lu)\,,\quad
R_1 = \frac{\sinh\frac{(1-2u)L}{2}}{\sinh\frac{L}{2}}\,.
\label{3:R}
\end{equation}
This gives us the $\beta_0^2$ parts of $\eta_{V,A}^{(3)}$
as exact functions of $r$.

Collecting all this together, we obtain
\begin{eqnarray}
&&\frac{\eta_{V,A}^{(3)}}{C_F} =
\left[ L \coth\frac{L}{2} \left( L^2 + 12 \zeta_2 + \eo{41/6}{39/2} \right)
\right.
\nonumber\\
&&\quad\left.{}
- \eo{6}{8} L^2 - \eo{24}{32} \zeta_2 - \eo{41/3}{326/9} \right] \beta_0^2
\nonumber\\
&&{} + \left( \frac{9}{2} C_F^2 + 12 C_F T_F + \frac{16}{3} T_F^2 \right) L^3
\nonumber\\
&&{} + \left[
\left( \eo{3}{15} \frac{C_F}{2} + \eo{4}{20} \frac{T_F}{3} \right) \beta_0
+ 3 \left( 16 \zeta_2 - \frac{1}{2} \eo{23}{29} \right) C_F^2
\right.
\nonumber\\
&&\quad{}
- 3 ( 4 \zeta_2 + 1 ) C_F C_A
+ \frac{16}{3} \left( 8 \zeta_2 - \eo{8}{11} \right) C_F T_F
\nonumber\\
&&\quad\left.{}
- \frac{4}{3} ( 8 \zeta_2 - 13 ) C_A T_F
- \eo{32/3}{128/9} T_F^2 \right] L^2
\nonumber\\
&&{} + \left[
\left( - \eo{48}{176/3} \pi^2 \log 2 + \eo{12}{28} \zeta_3
+ \eo{640}{896} \frac{\zeta_2}{3} - \eo{79/12}{213/4} \right) C_F \beta_0
\right.
\nonumber\\
&&\quad{}
+ \left( \eo{24}{88/3} \pi^2 \log 2 - \eo{54}{62} \zeta_3
- \eo{226}{322} \frac{\zeta_2}{3} + \frac{1}{3} \eo{56}{122} \right)
C_A \beta_0
\nonumber\\
&&\quad{}
+ \left( \eo{22}{2/3} \zeta_2 - \frac{1}{9} \eo{355}{179} \right) T_F \beta_0
\nonumber\\
&&\quad{}
+ \left( - \eo{48}{32} \pi^2 \log 2 + 80 \zeta_4 + \eo{36}{12} \zeta_3
+ \eo{14}{-146} \zeta_2 + \frac{1}{2} \eo{241}{385} \right) C_F^2
\nonumber\\
&&\quad{}
+ \left( \eo{24}{16} \pi^2 \log 2 + 16 \zeta_4 + \eo{6}{18} \zeta_3
+ \eo{47}{215} \frac{\zeta_2}{3} - \frac{1}{12} \eo{1313}{1769} \right)
C_F C_A
\nonumber\\
&&\quad{}
+ \left( 24 \zeta_4 - 132 \zeta_3 + \frac{16}{3} \zeta_2
+ \frac{1}{2} \eo{27}{-1} \right) C_A^2
\nonumber\\
&&\quad{}
+ \left( - \eo{128/3}{256/9} \pi^2 \log 2 - \eo{0}{64} \frac{\zeta_3}{3}
+ \eo{393}{-343} \frac{\zeta_2}{9} + \eo{481/2}{4273/18} \right) C_F T_F
\nonumber\\
&&\quad{}
+ \left( \eo{64/3}{128/9} \pi^2 \log 2 - \eo{96}{256/3} \zeta_3
+ \eo{-8/3}{40} \zeta_2 - \eo{152/3}{80} \right) C_A T_F
\nonumber\\
&&\quad\left.{}
- \frac{4}{9} \left( \eo{222}{158} \zeta_2 - \eo{1123/3}{265} \right) T_F^2
\right] L
\nonumber\\
&&{} + \frac{\eta^{(3)\prime}_{V,A}}{C_F}\,.
\label{3:eta}
\end{eqnarray}
Non-logarithmic terms of orders $\mathcal{O}(\beta_0)$ and $\mathcal{O}(1)$
are only known~\cite{AC:04} in the equal-mass limit $r=1$:
$\eta^{(3)\prime}_V=0$ and
\begin{eqnarray}
&&\frac{\eta^{(3)\prime}_A}{C_F} =
\biggl( \frac{256}{3} a_4 + \frac{32}{9} \log^4 2
+ \frac{64}{9} \pi^2 \log^2 2 - \frac{160}{9} \pi^2 \log 2
\nonumber\\
&&\qquad{}
- \frac{440}{3} \zeta_4 + \frac{184}{3} \zeta_3 + \frac{448}{3} \zeta_2
- \frac{2293}{18} \biggr) C_F \beta_0
\nonumber\\
&&{} + \biggl( - \frac{128}{3} a_4 - \frac{16}{9} \log^4 2
- \frac{32}{9} \pi^2 \log^2 2 + \frac{80}{9} \pi^2 \log 2
\nonumber\\
&&\qquad{}
+ \frac{220}{3} \zeta_4 - \frac{152}{3} \zeta_3 - \frac{20}{3} \zeta_2
- \frac{496}{9} \biggr) C_A \beta_0
\nonumber\\
&&{} + \left( - \frac{512}{3} \zeta_2 + \frac{7000}{27} \right) T_F \beta_0
\nonumber\\
&&{} + \biggl( - \frac{2560}{9} a_4 - \frac{320}{27} \log^4 2
+ \frac{320}{27} \pi^2 \log^2 2 - \frac{224}{3} \pi^2 \log 2
\nonumber\\
&&\qquad{}
- \frac{1280}{3} \zeta_5 + 224 \zeta_2 \zeta_3 - \frac{560}{3} \zeta_4
+ \frac{2560}{9} \zeta_3 + \frac{2480}{9} \zeta_2 - \frac{1141}{9}
\biggr) C_F^2
\nonumber\\
&&{}+ \biggl( \frac{512}{9} a_4 + \frac{64}{27} \log^4 2
- \frac{64}{27} \pi^2 \log^2 2 - \frac{560}{9} \pi^2 \log 2
\nonumber\\
&&\qquad{}
+ \frac{1040}{3} \zeta_5 + \frac{64}{3} \zeta_2 \zeta_3
+ \frac{656}{3} \zeta_4 - \frac{3920}{9} \zeta_3 + \frac{676}{9} \zeta_2
+ \frac{4777}{18} \biggr) C_F C_A
\nonumber\\
&&{}+ \biggl( \frac{128}{3} a_4 + \frac{16}{9} \log^4 2
- \frac{16}{9} \pi^2 \log^2 2 + \frac{448}{9} \pi^2 \log 2
\nonumber\\
&&\qquad{}
- \frac{320}{3} \zeta_5 + \frac{88}{3} \zeta_2 \zeta_3
- \frac{548}{3} \zeta_4 + \frac{844}{3} \zeta_3 - \frac{1658}{9} \zeta_2
- \frac{322}{9} \biggr) C_A^2
\nonumber\\
&&{}+ \biggl( \frac{13312}{9} a_4 + \frac{1664}{27} \log^4 2
- \frac{1280}{27} \pi^2 \log^2 2 + \frac{1408}{27} \pi^2 \log 2
\nonumber\\
&&\qquad{}
+ \frac{1696}{9} \zeta_4 - \frac{9152}{27} \zeta_3
- \frac{39296}{81} \zeta_2 + \frac{16958}{81}
\biggr) C_F T_F
\nonumber\\
&&{}+ \biggl( - \frac{9728}{9} a_4 - \frac{1216}{27} \log^4 2
+ \frac{1024}{27} \pi^2 \log^2 2 + \frac{17728}{27} \pi^2 \log 2
- 160 \zeta_5
\nonumber\\
&&\qquad{}
+ \frac{448}{3} \zeta_2 \zeta_3 - \frac{4208}{9} \zeta_4
- \frac{17312}{27} \zeta_3 - \frac{217232}{81} \zeta_2 + \frac{112064}{81}
\biggr) C_A T_F
\nonumber\\
&&{}+ \left( \frac{2048}{3} \zeta_3 - \frac{640}{3} \zeta_2
- \frac{39920}{81} \right) T_F^2\,,
\label{3:etap}
\end{eqnarray}
where $a_4=\mathop{\mathrm{Li}}\nolimits_4(1/2)$.
Of course, for the real value of $r$ this is an estimate only.
Assigning a $\pm100\%$ theoretical error to it,
we obtain a numerical estimate of the three-loop contribution
to~(\ref{Intro:1step}):
\begin{eqnarray}
&&\left[ \left( 1.81 + 2.89 \frac{r}{1-r} \right) L^3
- \eo{8.15}{9.76} L^2
+ \left( \eo{27.6}{45.6} + \eo{76.9}{113.5} \frac{r}{1-r} \right) L \right.
\nonumber\\
&&\left.{} - \eo{73.0}{124.6} \pm 3.9 \right]
\left(\frac{\alpha_s^{(5)}(\mu_0)}{\pi}\right)^3
\label{3:1step}
\end{eqnarray}
(we used $\coth(L/2)=(1+r)/(1-r)$).

\section{Numerical results}
\label{S:Conc}

The program RunDec~\cite{CKS:00} was used for numerical estimates:
\begin{eqnarray}
&&m_b \approx 4.7\;\mathrm{GeV}\,,\quad
m_c \approx 1.6\;\mathrm{GeV}\,,\quad
r \approx 0.34\,,\quad
L \approx 1.08\,,
\nonumber\\
&&\alpha_s(m_b) \approx 0.216\,,\quad
x \approx 1.56\,,\quad
\alpha_s^{(5)}(\mu_0) \approx 0.258\,.
\label{rundec}
\end{eqnarray}
With these numbers, the two-loop results~\cite{CM:97} give
\begin{eqnarray}
&&\eta_V \approx 1 + 0.1899 \frac{\alpha_s^{(5)}}{\pi}
+ 0.5416 \left(\frac{\alpha_s^{(5)}}{\pi}\right)^2
\approx 1.019\,,
\nonumber\\
&&\eta_A \approx 1 - 0.4768 \frac{\alpha_s^{(5)}}{\pi}
- 1.3813 \left(\frac{\alpha_s^{(5)}}{\pi}\right)^2
\approx 0.952\,.
\end{eqnarray}

To combine the results of the two-loop calculation~\cite{CM:97}
with our resummation,
we subtract from~\cite{CM:97} its $\mathcal{O}(r^0)$ term
and add the resummed result~(\ref{NNL:num}) instead.
This results in the corrections
\begin{equation}
\Delta\eta_V \approx 0.021\,,\quad
\Delta\eta_A \approx 0.029\,,
\label{Conc:corr1}
\end{equation}
to be added to the results of~\cite{CM:97}.
If we also add the $\beta_0^2$ term~(\ref{NNL:NNNL})
of the N$^3$L logarithmic correction,
these numbers change significantly:
\begin{equation}
\Delta\eta_V \approx -0.049\,,\quad
\Delta\eta_A \approx -0.059\,.
\label{Conc:corr2}
\end{equation}

It is more instructive to discuss the $\alpha_s^3$ terms~(\ref{3:1step})
in~(\ref{Intro:1step}).
They give
\begin{eqnarray}
&&\Delta\eta_V \approx - ( 9.7 \pm 3.9)
\left(\frac{\alpha_s^{(5)}(\mu_0)}{\pi}\right)^3
\approx - 0.005 \pm 0.002\,,
\nonumber\\
&&\Delta\eta_A \approx - (19.6 \pm 3.9)
\left(\frac{\alpha_s^{(5)}(\mu_0)}{\pi}\right)^3
\approx - 0.011 \pm 0.002\,.
\label{Conc:corr3}
\end{eqnarray}
The $\beta_0^2$ terms in~(\ref{3:eta}),
i.e., the naive nonabelianization prescription~\cite{BG:95},
give the coefficients of $(\alpha_s/\pi)^3$ in~(\ref{Conc:corr3})
equal to $0.9$ and $-14.0$ for $\eta_V$ and $\eta_A$;
in the first case, even the sign is wrong.
The authors of~\cite{AC:04} estimated these coefficients
as $0\pm3.9$ and $-11.1\pm3.9$,
on the basis of their equal-mass calculation
(our error estimate is the same as their one).
Our results differ from their ones by about two estimated errors.
The reason for this difference is the terms with powers of $L$
which have been investigated in the present paper.

Our final results
\begin{equation}
\eta_V = 1.014 \pm 0.002\,,\quad
\eta_A = 0.941 \pm 0.002
\label{final}
\end{equation}
have theoretical errors of about $0.2\%$;
the central values differ from the previous results~\cite{AC:04}
by about two estimated errors.

\end{document}